\documentstyle[12pt]{article}
\input{epsf}
\renewcommand{\theequation}{\arabic{section}.\arabic{equation}}
\newcommand{\be}{\begin{eqnarray}}
\newcommand{\ee}{\end{eqnarray}}
\newcommand{\ba}{\begin{array}}
\newcommand{\ea}{\end{array}}
\newcommand{\tr}{\mbox{\rm tr}}
\newcommand{\partialboth}{\stackrel{\leftrightarrow}{\nabla}}

\begin{document}
%
% BEGIN TITLEPAGE
%
\rightline{hep-ph/9809483}
\rightline{RUB-TPII-14/98}
\vspace{0.5cm}
\begin{center}
{\Large
Hard exclusive electroproduction of two pions and their resonances}
\\
\vspace{0.5cm}
{\large  M.V.\ Polyakov}

\vspace{0.2cm}
{Petersburg Nuclear Physics
Institute, Gatchina, St.\ Petersburg 188350, Russia}\\
and\\
Institut f\"ur Theoretische Physik II,
Ruhr--Universit\"at Bochum, D--44780 Bochum, Germany

%
%\date{\today}
%
\end{center}
\vspace{0.5cm}

\begin{abstract}
We study the hard exclusive production of two pions in the virtual
photon fragmentation region with various invariant masses including the
resonance region. The amplitude is expressed in terms of two-pion light
cone distribution amplitudes ($2\pi$DA's). We derive dispersion
relations for these amplitudes, which enables us to fix them completely
in terms of $\pi\pi$ scattering phases and a few low-energy subtraction
constants determind by the effective chiral lagrangian. Quantitative
estimates of the resonance as well $\pi\pi$ background DA's at low
normalization point are made. We also prove certain new soft pion
theorem relating two-pion DA's to the one-pion DA.
Crossing relations between $2\pi$DA's and parton distributions
in a pion are discussed.

\noindent
We demontrate that by studying the {\it shape} of the $\pi\pi$
mass spectra (not absolute cross section!) in a diffractive
electroproduction one can extract the deviation of the meson
($\pi,\rho,$~etc.) wave functions from their asymptotic form
$6 z(1-z)$ and hence to get important information about the structure
of mesons. Also we discuss how the moments of
quark distributions in a pion can be measured in the hard diffraction.
We suggest (alternative to S\"oding's) parametrization
of $\pi\pi$ spectra which is suitable for large photon virtuality.

\end{abstract}
\vspace{0.2cm}
%
%\pacs{PACS: 12.38.Lg, 13.60.Fz, 13.60.Le}
%
% BEGIN TEXT
%

\newpage
\tableofcontents

\newpage

\section{Introduction}
\setcounter{equation}{0}

Recently it became possible to measure with high precision two pion
hard
diffractive production in the deeply virtual photon fragmentation
region. The paper contains a  detailed study of the ``upper blob"
of this process, more precisely, of the two-pion distribution
amplitude.

It has been realized recently \cite{CFS,Rad,Ji} that a number of
hard exclusive processes of the type:

\be
\gamma_L^\ast(q)+ T(p) \to F(q') + T'(p')\;,
\label{proc1}
\ee
at large $Q^2$ with $t=(p-p')^2$, $x_{Bj}=Q^2/2(pq)$ and
$q'^2=M_F^2$,  fixed are amenable to QCD description. The
factorization theorem \cite{CFS} asserts that the amplitude
of the process (\ref{proc1}) has the following form:

\begin{eqnarray}
   &&
   \sum _{i,j} \int _{0}^{1}dz  \int dx\,
   f_{i/T}^{T'}(x ,x -x_{Bj};t,\mu ) \,
   H_{ij}(Q^2 x/x_{Bj},Q^{2},z,\mu )
   \, \Phi^F_{j}(z,\mu )
\nonumber\\
&&
   + \mbox{power-suppressed corrections} ,
\label{factorization}
\end{eqnarray}
where $f_{i/T}^{T'}$ is $T\to T'$ skewed parton distribution
\cite{Dittes,Dmuller,CFS,Rad,Ji} (for a review see \cite{JiReview}),
$\Phi^F_{j}(z,\mu )$ is the distribution amplitude of the hadronic
state $F$ (not necessarily one particle state), and $H_{ij}$ is a hard
part computable in pQCD as series in $\alpha_s(Q^2)$. Here we shall
study general properties of the distribution amplitudes
$\Phi^F_{j}(z,\mu )$ when the final hadronic state $F$ is a two pion
state ($F=\pi\pi$).

The $2\pi$DA's were introduced recently in \cite{Ter} in the context of
the QCD description of the process $\gamma^\ast \gamma \to 2 \pi$, first
model estimate of the corresponding DA at the
two-pion invariant mass below
the resonance region was done in \cite{PW98}. In the present paper we
discuss how to describe the resonance region.
We shall see that calculations of the two-pion DA in low
energy region \cite{PW98} and the knowledge of $\pi\pi$ scattering
phases are sufficient to determine $2\pi$DA's
in a wide range of two-pion invariant masses.

For definiteness we shall  consider  hard electroproduction
of two pions off the nucleon:

\be
\gamma_L^\ast + N \to 2\pi +T'\; ,
\label{proc2}
\ee
where the final state $T'$ can be $N,\Delta, \pi N,\ $etc.
The usage of $2\pi$DA's to describe the process (\ref{proc2})
has several advantages comparing to the case of
single resonance DA's
($\rho\, , \rho'\, ,$~etc.):
\begin{itemize}
\item
The $2\pi$DA's are the most general objects to describe the
form of the $\pi\pi$ mass
spectra in diffractive production of pions.
We shall see that by studying the shape
of the two-pion mass spectra
(not the absolute cross section!)
in diffractive production
of two pions  one can extract
the information about the deviation of the meson ($\pi,\rho$) DA's from
their asymptotic form, $i.e.$ essentially non-perturbative information
about the structure of mesons.
\item
The usage of $2\pi$DA's is anyhow necessary in studies of channels with
strong nonresonant background, for example,  in the case of hard
electroproduction of two neutral pions.
\item
By crossing symmetry the $2\pi$DA's are related to skewed parton
distribution (SPD's) in a pion (see section~\ref{crossing}). Were
$2\pi$DA's measured one could get information about SPD's in a pion.
\item From theoretical point of view $2\pi$DA's are necessary to define
DA's of unstable resonances ($\rho, \rho', f_2, \rho_3,\, $etc.).
\end{itemize}

The plan of this paper is as follows.  Section~\ref{odin} contains
definitions and general properties of two-pion distribution
amplitudes.
In Section~\ref{softpi} we prove soft pion theorems
for the $2\pi$DA's. These theorems relate the $2\pi$DA's to the
DA's of a single pion in a limit when one of the produced pions is
soft.
In section~\ref{crossing} we derive relations between $2\pi$DA's and
quark distributions (skewed and not) in a pion. These relations
can be, in principle, used for measurements of quark distributions
in a pion in hard exclusive reactions.
Also they constrain the $2\pi$DA's by already known
parton distributions in a pion.
Section~\ref{tri} is devoted to derivation of the dispersion
relations for the $2\pi$DA's. These dispersion relations allow us to
express the $2\pi$DA's in a wide range of invariant two-pion masses
in terms of (known) pion--pion phase shifts
and a few low-energy subtraction constants.
In subsection~\ref{reson} we explain how to extract
resonance DA's from $2\pi$DA's. In section~\ref{instantons} we
estimate the low--energy subtraction constants in the
instanton model of the QCD
vacuum, also results for $\rho$-meson DA's are presented. In
section~\ref{hard} we discuss the applications and implications of the
$2\pi$DA's to the hard electroproduction of pions.
Conclusions and an
outlook are given in section~\ref{sec_conclusions}.

\section{Two-pion light-cone distribution amplitudes.}
\setcounter{equation}{0}
\label{odin}

We shall consider the following twist-2 chirally even and odd two--pion
distribution amplitudes (2$\pi$DA's)
\be
\nonumber
\Phi^{ab}_{\|} (z, \zeta, w^2 )
&=&
 \frac{1}{4\pi} \int dx^- e^{-\frac{i}{2}z P^+ x^-}\\
&\times&
{}_{out}\langle
\pi^a(p_1) \, \pi^b(p_2) \, | \, \bar\psi(x) \; \hat n \, T \; \psi(0)\,
|0\rangle \Bigr|_{x^+=x_\perp=0} ,
\label{definition1}
\ee

\be
\nonumber
\Phi^{ab}_{\perp} (z, \zeta, w^2 ) &=&
\frac{i\,f_{2\pi}^\perp}{4\pi w^2} \int dx^- e^{-\frac{i}{2}z
P^+ x^-} \\
&\times&{}_{out}\langle \pi^a(p_1) \, \pi^b(p_2) \, | \, \bar\psi(x)
\; \sigma_{\mu\nu}n^\mu P^\nu \, T \; \psi(0)\, |0\rangle
\Bigr|_{x^+=x_\perp=0} ,
\label{definition2}
\ee

Here, $n$ is a light--like vector ($n^2 = 0$), which we choose to be
$n_\mu=(1,0,0,1)$. For any vector, $V$, the ``plus'' light--cone
coordinate is defined as $V^+ \equiv (n\cdot V) = V^0 + V^3$; the
``minus'' component as $V^- = V^0 - V^3$. The outgoing pions have
momenta $p_1, p_2$, and $P \equiv p_1 + p_2$ is the total momentum of
the final state, with $P^2=w^2$.
Finally, $T$ is a flavor matrix ($T = 1/2$ for the
isosinglet, $T = \tau^3/2$ for the isovector $2\pi$DA).
[In the above equations
the path--ordered exponential of the gauge field, required by
gauge invariance is always assumed].
The constant
of dimension of mass $f_{2\pi}^\perp$ is defined as matrix element of
local operator:
\be
\lim_{w^2\to 0}
\langle \pi^a(p_1) \, \pi^b(p_2) \, | \, \bar\psi(0) \;
\sigma^{\mu\nu} \; \frac{\tau^3}{2} \psi(0)\, |0\rangle=
\varepsilon^{ab3}\frac{2\, i}{f_{2\pi}^\perp}(p_1^\mu p_2^\nu-p_1^\nu
p_2^\mu).
\label{f2pi}
\ee

The generalized distribution amplitudes,
eqs.~(\ref{definition1},\ref{definition2}), depend
on the following kinematical variables: the quark momentum fraction with
respect to the total momentum of the two--pion state, $z$; the variable
$\zeta \equiv p_1^+ / P^+$ characterizing the distribution of
longitudinal momentum between the two pions, and the invariant mass
(c.m.\ energy), $w^2 = P^2$.
\par
In what follows we shall work in the reference frame where
$P_\perp = 0$.  In this frame
\be
P^-&=&\frac{w^2}{P^+}, \qquad p_1^- \; = \; \frac{w^2(1-\zeta)}{P^+},
\qquad
p_\perp^2 \; \equiv \; p_{1\perp}^2 \; = \; p_{2\perp}^2 \; = \;
w^2\zeta(1-\zeta)-m_\pi^2.
\hspace{.5cm}
\ee
From these relations one obtains the following kinematical constraint:
\be
\zeta(1-\zeta)\geq \frac{m_\pi^2}{w^2} .
\label{con}
\ee

The isospin decomposition of the $2\pi$DA's
eqs.~(\ref{definition1},\ref{definition2}) has the form\footnote{
We do not specify the indices ($\|$ and $\perp$) if
an equation holds for
both chirally even and odd DA's}:

\be
\Phi^{ab}=\delta^{ab}\tr(T)\Phi^{I=0} +\frac12 \tr([\tau^a,\tau^b] T)
\Phi^{I=1}.
\ee
From the C-parity one can easily derive the following symmetry
properties of the
isoscalar ($I=0$) and isovector ($I=1$) parts of $2\pi$DA's
eqs.~(\ref{definition1},\ref{definition2}):

\be
\nonumber
\Phi^{I=0}(z,\zeta,w^2)&=&
-\Phi^{I=0}(1-z,\zeta,w^2)=\Phi^{I=0}(z,1-\zeta,w^2),\\
\Phi^{I=1}(z,\zeta,w^2)&=&\Phi^{I=1}(1-z,\zeta,w^2)=
-\Phi^{I=1}(z,1-\zeta,w^2)\; .
\label{reflection}
\ee

The isospin one parts of $2\pi$DA's
eqs.~(\ref{definition1},\ref{definition2}) are normalized as follows:
\be
\nonumber
\int_0^1 dz \Phi_{\|}^{I=1} (z, \zeta, w^2 )=(2\zeta-1) F_\pi(w^2)\; ,\\
\int_0^1 dz \Phi_{\perp}^{I=1} (z, \zeta, w^2 )=(2\zeta-1) F_{t}(w^2)\; ,
\label{norm}
\ee
where $F_\pi(w^2)$ is the pion e.m. form factor in the time-like region
($F_{\pi}(0)=1$) and $F_{t}(w^2)$ is the ``tensor" form factor of the
pion normalized by $F_{t}(0)=1$.

Let us decompose $2\pi$DA's in eigenfunctions of the ERBL \cite{ERBL}
evolution equation (Gegenbauer polynomials $C_n^{3/2}(2 z-1)$) and
in partial waves of produced pions\footnote{In the formulae below
we, for simplicity, neglect threshold effects. How to take them into
account is discussed in the Appendix.} (Gegenbauer polynomials
$C_l^{1/2}(2 \zeta-1)$, see below).  Generically the decomposition (for
both isoscalar and isovector DA's) has the form:

\be
\Phi (z, \zeta, w^2 )=6z(1-z)\sum_{n=0}^{\infty}
\sum_{l=0}^{n+1} B_{nl}(w^2) C_n^{3/2}(2 z-1)C_l^{1/2}(2
\zeta-1).  \label{razhl} \ee
Using the symmetry properties
(\ref{reflection}) we see that the index $n$ goes over even (odd) and
index $l$ goes  over odd (even) numbers for isovector (isoscalar)
$2\pi$DA's.  The normalization conditions (\ref{norm}) correspond
to $B_{01}^\|(w^2)=F_\pi(w^2)$ and $B_{01}^\perp(0)=1$.

The Gegenbauer polynomials $C_n^{3/2}(2 z-1)$ are eigenfunctions of the
ERBL \cite{ERBL} evolution equation and hence the coefficients
$B_{nl}$ are renormalized multiplicatively (for even $n$ and odd $l$):

\be
B_{nl}(w^2;\mu)=B_{nl}(w^2;\mu_0)
\biggl(
\frac{\alpha_s(\mu)}{\alpha_s(\mu_0)}
\biggr)^{(\gamma_n-\gamma_0)/(2\beta_0)}\;,
\label{renorm}
 \ee
where $\beta_0=11-2/3 N_f$ and the one loop anomalous dimensions are
\cite{VY}:
\be
\nonumber
\gamma_n^{\|}&=&\frac{8}{3}
\Bigl(1-\frac{2}{(1+n)(2+n)}+4\sum_{k=2}^{n+1}\frac{1}{k} \Bigr) \;,\\
\nonumber
\gamma_n{^\perp}&=&\frac{8}{3}
\Bigl (1+4\sum_{k=2}^{n+1}\frac{1}{k} \Bigr) \;.
 \ee

From the decomposition (\ref{razhl}) and eq.~(\ref{renorm}) we can
immediately
make a simple prediction for the ratio of the
hard $P-$ and $F-$ waves production amplitudes of
pions in the reaction $\gamma^\ast p\to 2 \pi p$
at large virtuality of the incident photon
and fixed $w^2$ and Bjorken $x$:

\be
\frac{F{\rm -wave\ amplitude}}{P{\rm -wave\ amplitude}}\sim
\frac{1}{\log(Q^2)^{50/(99-6 N_f)}}\; ,
\ee
or generically for the $2 k+1$ wave:
\be
\frac{(2k+1) {\rm -wave\ amplitude}}{P {\rm -wave\ amplitude}}\sim
\frac{1}{\log(Q^2)^{\gamma_{2k}^\|/(2\beta_0)}}\; .
\ee

\section{Production of soft pions}
\setcounter{equation}{0}
\label{softpi}

In this section we prove the low-energy theorem for
the isovector $\Phi_{\|}^{I=1} (z, \zeta, w^2 )$ relating this $2\pi$DA
to the single pion DA $\varphi_\pi(z)$ \cite{doklad}.  The latter is
defined as:

\be
\langle \pi^0 (P ) | \bar \psi (x) \hat n \gamma_5
\frac{\tau^3}{2}  \psi (0)
| 0 \rangle
= i  f_\pi P^+
\int_0^1 dz \; e^{i z P\cdot x} \varphi_\pi (z) .
\label{phi_pion}
\ee
Here $f_\pi=93$~MeV is the pion decay constant.

Let us consider the matrix element
$
{}_{out}\langle
\pi^a(p_1) \, \pi^b(p_2) \, | \, \bar\psi(x) \; \hat n \,
\frac{\tau^3}{2} \; \psi(0)\,
|0\rangle
$ entering the definition of the chirally even isovector $2\pi$DA
eq.~(\ref{definition1}) and consider the kinematical limit
when the 4-momentum of one of the pions goes to zero. This limit
corresponds to $w^2\to 0$ and $\zeta\to 1$ (or $\zeta\to 0$).
Using the LSZ reduction formula we can write (for definiteness we take
$p_2^\mu\to 0$ which corresponds to $w^2\to 0$ and $\zeta\to 1$):

\be
\nonumber
{}_{out}\langle
\pi^a(p_1) \, \pi^b(p_2) \, | \, \bar\psi(x) \; \hat n \,
\frac{\tau^3}{2} \; \psi(0)\,
|0\rangle &=&
i\int d^4 y e^{i(p_2\cdot y)}(\partial_y^2+m_\pi^2)\\
&\times&\langle
\pi^a(p_1) |T\biggl\{ \phi_\pi^b(y)\,\bar\psi(x) \; \hat n \,
\frac{\tau^3}{2} \; \psi(0)\biggr\}
|0\rangle,
\label{f1}
\ee
where $\phi_\pi^b(y)$ is a pion interpolating field which, owing to
PCAC, can be related to the divergence of the axial current
$\phi_\pi^b(y)=\frac{1}{f_\pi m_\pi^2}\partial_\mu J_5^{\mu b}(y) $.
Substituting this into eq.~(\ref{f1}) and taking the
limit $p_2^\mu\to 0$
we can rewrite r.h.s. of eq.~(\ref{f1}) in the form:

\be
\nonumber
&&\frac{i}{f_\pi}\int d^4 y
\langle
\pi^a(P) |T\biggl\{ \partial_\mu J_5^{\mu b}(y)\,\bar\psi(x) \; \hat n \,
\frac{\tau^3}{2} \; \psi(0)\biggr\}
|0\rangle=\\
&&\frac{1}{i f_\pi}\int d^4 y
\Bigl\{
\delta(y^0)
\langle
\pi^a(P) | \bar\psi(x) \; \hat n \,
\frac{\tau^3}{2} \; [J_5^{0 b}(y),\,\psi(0)]
|0\rangle\\
\nonumber
&&+
\delta(y^0-x^0)
\langle
\pi^a(P) | [J_5^{0 b}(y),\,\bar\psi(x)] \; \hat n \,
\frac{\tau^3}{2} \; \psi(0)
|0\rangle
\Bigr\}.
\label{f2}
\ee
The equal time commutators entering the above equation
are simply the global axial transformations of the quark fields:

\be
\nonumber
\int d^4 y\, \delta(y^0) [J_5^{0 b}(y),\,\psi(0)]&=&
i\frac{\tau^b}{2}\gamma_5 \psi(0) \; , \\
\int d^4 y \, \delta(y^0-x^0) [J_5^{0 b}(y),\,\bar \psi(x)]&=&
i\bar\psi(x) \frac{\tau^b}{2}\gamma_5 \; .
\ee
Substituting this result into eq.~(\ref{f2}) we get the following
soft pion theorem:

\be
\lim_{p_2\to 0}{}_{out}\langle
\pi^a(p_1) \, \pi^b(p_2) \, | \, \bar\psi(x) \; \hat n \,
\frac{\tau^3}{2} \; \psi(0)\,
|0\rangle =  \frac{i \varepsilon^{3 b c}}{f_\pi}
\langle \pi^a (P ) | \bar \psi (x) \hat n \gamma_5
\frac{\tau^c}{2}  \psi (0)
| 0 \rangle \, .
\label{let}
\ee
Applying this soft pion theorem to the matrix elements entering
the definition of the chirally even
$2\pi$DA eq.~(\ref{definition1}) and comparing the resulting expression
with definition of the pion DA, eq.~(\ref{phi_pion}), we get the
following relation between the pion DA and the isovector chirally even
$2\pi$DA:

\be
\Phi_\|^{I=1}(z,\zeta=1,w^2=0)=
-\Phi_\|^{I=1}(z,\zeta=0,w^2=0)=\varphi_\pi(z)\, .
\label{letwf}
\ee
The analogous theorem for the isoscalar part of the $2\pi$DA's has the
form:
 \be \Phi_\|^{I=0}(z,\zeta=1,w^2=0)=
\Phi_\|^{I=0}(z,\zeta=0,w^2=0)=0\; .
\label{letwf0}
\ee

Some comments are in order here. The derivation of the soft pion
theorems presented here can be easily extended to other cases
when we have an emission of soft pions. We should only keep in mind that
for the case when we have nucleons in the $out$ state (a typical
example is a generalized skewed parton distributions, see
\cite{FPS,MVPB98}) the
singularities related to the emission of soft pions from the nucleon
legs must be taken into account.

The soft pion theorem eq.~(\ref{letwf}) allows us to relate
the coefficients $B_{nl}^\|(w^2)$ (see eq.~(\ref{razhl}))
and the coefficients in the expansion of the pion DA in
the Gegenbauer
polynomials
\be \varphi_\pi(z)=6 z(1-z)\biggl ( 1+\sum_{n={\rm even}}
a_n^{(\pi)} C_n^{3/2}(2 z-1) \biggr).
\label{razhlpi}
\ee
The relation has the form:

\be
a_n^{(\pi)}=\sum_{l=1}^{n+1} B_{nl}^{\|(I=1)}(0)\; .
\label{letcoef}
\ee
The soft pion theorem for the isoscalar chirally even $2\pi$DA
eq.~(\ref{letwf0}) implies that
\be
\sum_{l=0}^{n+1} B_{nl}^{\|(I=0)}(0)=0\; .
\label{let0}
\ee

\section{Crossing relations between $2\pi$DA's and quark distributions
in a pion}
\setcounter{equation}{0}
\label{crossing}

An interesting feature of the pion is the fact that the $2\pi$DA's
are related to the skewed parton distribution in a pion by crossing.
On one hand, the crossing relations derived in this section
\footnote{Part of the material of this section was presented already
in recent preprints \cite{Dubrovnik,PW99}.}
allow to use the information about quark distributions in a pion to
constrain two-pion distribution amplitudes. On other hand, the
measurements of $2\pi$DA's in hard exclusive reactions can be used as a
source of information on skewed quark distributions in a pion.
Additionally the crossing relation are helpfull in modelling of
skewed parton distributions.

The matrix element entering in the definition of $2\pi$DA (\ref{definition1})
by crossing is related to the matrix element in definition
of skewed distribution:

\be
&&\int \frac{d\lambda }{2\pi }
\exp[i \lambda \tau n\cdot(p+p')]
\langle \pi^a(p^{\prime })|T \bigl\{
\bar \psi_{f'}
(-\lambda n/2){\hat n}
 \psi_f (\lambda n/2)\bigr\}
|\pi^b(p)\rangle = \\
\nonumber
&&\delta^{ab}\delta_{ff'} H^{I=0}(\tau,\xi,t) +
i\varepsilon^{abc}
\tau^c_{ff'} H^{I=1}(\tau,\xi,t)
\; ,
\label{spd-def}
\ee
where $\xi$ is a skewedness parameter defined as:
$\xi=-(n\cdot (p'-p))/(n\cdot(p'+p))$,  $t=(p'-p)^2$ and light-cone
vector $n^\mu$ introduced in section~\ref{odin}.

This allows to express the moments
of skewed parton distributions in terms of expansion coefficients
$B_{nl}$
in eq.~(\ref{razhl}):
\be
\int_{-1}^1 d\tau \tau^{N-1} H^I(\tau,\xi,t)=
\sum_{n=0}^{N-1}
\sum_{l=0}^{n+1} B^I_{nl}(t) \xi^N\
C_l^{1/2}\biggl(\frac{1}{\xi}\biggr)
\int_{-1}^1 d x
\frac{3}{4}[1-x^2]\ x^{N-1} C_n^{3/2}(x) \, .
\label{relation}
\ee
Let us note that this expression satisfy the polynimiality
condition for moments of SPD's (see ref.~\cite{JiReview}).

In order to prove the relation (\ref{relation}) let us consider the
operator expression for the $N-$th moments of SPD. It is given by
off-forward matrix element of local operator of spin $N$ and twist-2:

\be
\int_{-1}^1 d \tau \tau^{N-1} H(\tau,\xi,t)=
\frac{1}{[(p'+p)^+]^N }
\langle p'| \bar \psi \gamma^+ (\partialboth^+)^{N-1} \psi |p\rangle\; ,
\label{cross1}
\ee
Analogously the $N-$th moments of $2\pi$DA (\ref{definition1})
is expressed in terms of the following matrix element of the same local
operator:

\be
\nonumber
\int_0^1 dz (2 z-1)^{N-1} \Phi^I(z,\zeta,w^2)=
\frac{1}{[(p'+p)^+]^N }
\langle p' p| \bar \psi \gamma^+ (\partialboth^+)^{N-1} \psi |0\rangle=\\
\sum_{n=0}^{N-1}
\sum_{l=0}^{n+1} B^I_{nl}((p'+p)^2)
C_l^{1/2}\biggl(\frac{(p'-p)^+}{(p'+ p)^+} \biggr)
\int_{-1}^1 d x
\frac{3}{4}[1-x^2]\ x^{N-1} C_n^{3/2}(x)
\; ,
\label{cross2}
\ee
where in the last equality we used the double expansion (\ref{razhl}).
The matrix elements in eq.~(\ref{cross2})
and in eq.~(\ref{cross1}) are related to each other by crossing:

\be
\nonumber
\frac{1}{[(p+p')^+]^N }
\langle p'| \bar \psi \gamma^+ (\partialboth^+)^{N-1} \psi |p\rangle=
\frac{1}{[(p'+p)^+]^N }
\langle -p, p'| \bar \psi \gamma^+ (\partialboth^+)^{N-1} \psi
|0\rangle= \\
\nonumber
\frac{[(p'-p)^+]^N}{[(p'+p)^+]^N} \sum_{n=0}^{N-1}
\sum_{l=0}^{n+1} B^I_{nl}((p'-p)^2)
C_l^{1/2}\biggl(\frac{(p'+p)^+}{(p'- p)^+} \biggr)
\int_{-1}^1 d x
\frac{3}{4}[1-x^2]\ x^{N-1} C_n^{3/2}(x) \,= \\
\sum_{n=0}^{N-1}
\sum_{l=0}^{n+1} B^I_{nl}(t) \xi^N\
C_l^{1/2}\biggl(\frac{1}{\xi}\biggr)
\int_{-1}^1 d x
\frac{3}{4}[1-x^2]\ x^{N-1} C_n^{3/2}(x) \, .
\ee
This is exactly relation (\ref{relation}). Let us note
that in the expression (\ref{relation}) the coefficients
$B_{nl}(t)$ enter at negative argument whereas in (\ref{razhl})
at positive argument. The corresponding analytical continuation
can be done with help of dispersion relations (\ref{dr}) (see the next
section).

If we take the forward limit in (\ref{relation}), we obtain the relations
between moments of quark distributions in a pion and coefficients
$B_{nl}(0)$:
\be
\nonumber
M_N^{(\pi)}&\equiv& \int_{0}^1 dx\ x^{N-1} (q_\pi(x)-\bar q_\pi(x))=
B^{I=1}_{N-1,N}(0) A_N \qquad \mbox{ for \ odd\ } N\, ,\\
M_N^{(\pi)}&\equiv& 2 \int_{0}^1 dx\ x^{N-1} (q_\pi(x)+\bar
q_\pi(x))= B^{I=0}_{N-1,N}(0) A_N \qquad \mbox{for \ even\ } N \, ,
\label{f2tp}
\ee
where $A_N$ are numerical coefficients ($e.g.$ $A_1=1, A_2=9/5,
A_3=6/7,A_4=5/3$, etc.) and $q_\pi(x)=u^{\pi^+}(x)$.
For example, from eq.~(\ref{f2tp}) we obtain that
$B^{I=1}_{01}(0)=M_1^{(\pi)}=1$ what corresponds to normalization
condition (\ref{norm}). Also it is easy to see that
$B^{I=0}_{12}(0)=5/9 M_2^{(\pi)}$ corresponds to normalization
for isoscalar $2\pi$DA
\footnote{To see this one has to use additionally soft
pion theorem (\ref{letwf0}) for isoscalar
$2\pi$DA}:

\be
\int_0^1 dz (2 z-1)\Phi^{I=0}(z, \zeta, w^2=0 )
=-2\ M_2^{\pi}\zeta(1-\zeta) \; ,
\label{norm1}
\ee
where $M_2^\pi$ is a momentum fraction carried by quarks in
a pion.

Using relations (\ref{f2tp}) and soft pion theorem
(\ref{letwf}) we can express the coefficient $B_{21}(0)$
[it describes the deviation of isovector $2\pi$DA from its
asymptotic form $\Phi^{I=1}_{as}(z,\zeta,w^2)=
6z(1-z)(2\zeta-1) F_\pi^{{\rm e.m.}}(w^2)$] as:

\be
B_{21}^{I=1}(0)=a_2^{(\pi)}-\frac 76 M_3^{(\pi)}\; .
\label{therelation}
\ee

{}From the expression for the moments of SPD (\ref{relation}) we can
formally construct the SPD itself as series:

\be
H^I(\tau,\xi,t)=
\sum_{n=0}^{\infty}\sum_{l=0}^{n+1}
\frac 34 (1-\frac{\tau^2}{\xi^2})\theta(|x|-|\xi|)\
B^I_{nl}(t)
C_n^{3/2}\bigl(\frac{\tau}{\xi} \bigr)
C_l^{1/2}\bigl(\frac{1}{\xi} \bigr)\; ,
\label{formal}
\ee
where each term in the sum has a support $-\xi < \tau<\xi$.
Let us note however that it does not imply that the SPD
$H(\tau,\xi,t)$ has the same support because generically the sum (\ref{formal})
is divergent at fixed $\tau$ and $\xi$. In the formal solution
(\ref{formal}) each individual term of the series gives the
contribution to the amplitude which behaves as
 $1/\xi^l$ at small $\xi$ what would imply the violation of
unitarity for the hard exclusive reactions. This ``disaster" is avoided
through the fact the series is divergent and one can not calculate the
asymptotic behaviour term by term. This situation is similar to
duality idea: the $entire$ skewed
and usual quark distributions  are given by infinite sum over
$t$--channel resonances. Of course, there might be ``non-dual"
or ``two component" way to compensate the bad behaviour of the
resonances by some equally bad behaviour of non-resonant background.
Such ``two component" model can be written as:
\be
\nonumber
H^I(\tau,\xi,t)&=&
\sum_{l=0}^{L_0}\sum_{n=l+1}^{\infty}
\frac 34 (1-\frac{\tau^2}{\xi^2})\theta(|x|-|\xi|)\
B^I_{nl}(t)
C_n^{3/2}\bigl(\frac{\tau}{\xi} \bigr)
C_l^{1/2}\bigl(\frac{1}{\xi} \bigr)\\
&+&H^I_{\rm non-resonant}(\tau,\xi,t)
\; , \label{twocomponent} \ee
where $L_0$ is the maximal spin of  resonances which are taken into
account and\\ $H^I_{\rm non-resonant}(\tau,\xi,t) $ is a non-resonant
background. The latter can be modelled in terms of double distributions
\cite{RadDD}, however if $L_0>0$ the non-resonant contribution has to
be oscillating function in order to avoid double counting in
calculations of low moments of SPD ($N\leq L_0$) which are saturated by
resonances in $t$-channel.

To summarize, the representation of SPD's (\ref{formal}) in terms of
only resonance exchages gives a new picture of quark distributions in
a pion which is $dual$ to the picture based on the concept of double
distributions \cite{RadDD}. It would be extremely interesting to build
a model where the duality relation
\be
\lim_{t, \xi\to 0}
\sum_{n=0}^{\infty}\sum_{l=0}^{n+1}
\frac 34 (1-\frac{\tau^2}{\xi^2})\theta(|x|-|\xi|)\
B^I_{nl}(t)
C_n^{3/2}\bigl(\frac{\tau}{\xi} \bigr)
C_l^{1/2}\bigl(\frac{1}{\xi} \bigr)= q_\pi(\tau)
\; ,
\ee
is realized explicitly.

At the end let us note that
from the formal solution (\ref{formal}) and soft pion theorems
(\ref{letwf},\ref{letwf0}) we can determine the shape of SPD's in a
pion in a limiting case: $\xi=1$ and $t=0$:

$$
H^{I=1}(\tau,\xi=1,t=0)= \frac 12 \varphi_\pi(\frac{\tau+1}{2})\; ,\ \
\
H^{I=0}(\tau,\xi=1,t=0)= 0\; , $$
where $\varphi_\pi(z)$ is a pion distribution amplitude.
These limiting relations impose strong constraints on the models of
skewed parton distributions. For example it is easy to see that the
model based on double distributions proposed in \cite{RadDD} does not
satisfy these constraints.

\section{Two-pion distribution amplitudes in the resonance region and
DA's of resonances}
\setcounter{equation}{0}
\label{tri}

Generally speaking, the $2\pi$ DA's are  complex functions
owing to the strong interaction of the produced pions.
Above the two-pion threshold $w^2=4 m_\pi^2$
the $2\pi$DA's develop an imaginary
part corresponding to the contribution of on-shell intermediate states
($2\pi$, $4\pi$, etc.). In the region $w^2<16 m_\pi^2$ the imaginary part
is related to the pion-pion scattering amplitude by the Watson theorem
\cite{Watson}. This relation can be written in the following form
(in this section we omit the indices $\|$ and $\perp$
but show the isospin index, since all formulae in this section apply to
both chirally even and odd $2\pi$DA's):

\be
\mbox{Im} \Phi^I(z,\zeta,w^2)=\int d({\rm phase\ space})
\Phi^I(z,\zeta',w^2)^\ast
A_{\pi\pi}^I(\zeta',\vec p_\perp'|\zeta,\vec p_\perp)\; ,
\label{w1}
\ee
where $A_{\pi\pi}^I(\zeta',\vec p_\perp'|\zeta,\vec p_\perp)$
is pion-pion scattering amplitude with isospin $I$ in the $s$-channel,
$\zeta'$ and $p_\perp'$ are the momenta of the intermediate pion,
the phase space integral in these variables has the form:

\be
\int d({\rm phase\ space}) =\frac{1}{64\pi^2}\int_0^1 d\zeta'
\int_0^{2\pi} d\theta_\perp,
\ee
where $\theta_\perp$ is the angle between $p_\perp$ and $p_\perp'$.
In order to find the form of the $\pi\pi$ scattering amplitude in the
light-cone variables we start with the usual partial wave expansion of
the pion-pion scattering amplitude:

\be
A_{\pi\pi}^I= 32\pi \sum_{l} (2l+1)
\sin[\delta_l^I(w^2)]e^{i\delta_l^I(w^2)} P_l[\cos(\theta_{cm})]\; ,
\label{pwe}
\ee
where $\theta_{cm}$ is the scattering angle in the centre of mass
system, it can be easily expressed in terms of light-cone variables. The
relation has the form (here for a massless pion for
simplicity)\footnote{Expressions for massive case see in Appendix.}:

\be
\cos(\theta_{cm})=1-2(\zeta+\zeta'-2\zeta\zeta')-
4\sqrt{\zeta(1-\zeta)\zeta'(1-\zeta')}\cos(\theta_\perp)\; .
\label{cmlc}
\ee

Substituting the
partial wave expansion (\ref{pwe}) into eq.~(\ref{w1})
and computing the integral over $\theta_\perp$ using the identity:

\be
\nonumber
\int_0^{2\pi} d\theta_\perp P_l\biggl[
1-2(\zeta+\zeta'-2\zeta\zeta')-
4\sqrt{\zeta(1-\zeta)\zeta'(1-\zeta')}\cos(\theta_\perp)\biggr]=\\
2\pi\,
C_l^{1/2}(2\zeta-1)C_l^{1/2}(2\zeta'-1)\;,
\label{additionth}
\ee
we get the following expression for the imaginary part of the
$2\pi$DA:

\be
\nonumber
\mbox{Im}\, \Phi^I(z,\zeta,w^2+i0)&=&\sum_l (2l+1)\int_0^1 d\zeta'
\Phi^I(z,\zeta',w^2)^\ast
\sin[\delta_l^I(w^2)]e^{i\delta_l^I(w^2)} \\
&\times&
C_l^{1/2}(2\zeta-1)C_l^{1/2}(2\zeta'-1) \; .
\label{qq}
\ee
If we now substitute the expansion of the $2\pi$DA in the Gegenbauer
polynomials (\ref{razhl})  in eq.~(\ref{qq})
we get the equation for the imaginary part of
the expansion coefficients $B_{nl}$:

\be
\mbox{Im}\, B_{nl}^I(w^2)=\sin[\delta_l^I(w^2)]e^{i\delta_l^I(w^2)}
B_{nl}^I(w^2)^\ast=\tan[\delta_l^I(w^2)]\mbox{Re}B_{nl}^I(w^2)\; .
\label{imbnl}
\ee
One sees immediately that the phases of the coefficients $B_{nl}^I$
are the same as the phase shifts $\delta_l^I$. Let us also note that
the equation (\ref{imbnl}) for the case $n=0, l=1$ and $I=1$
is identical to that obtained in \cite{guerro} for the pion electromagnetic
form factor, which is not surprising because of identity
$B_{01}(w^2)=F_\pi(w^2)$. Since the form of equation (\ref{imbnl})
is very similar to that for e.m. pion form factor one can analyze it
following ref.~\cite{guerro}.

Using  eq.~(\ref{imbnl}) one
 can write an $N$-subtracted dispersion relation for the
$B_{nl}^I(w^2)$:

\be
B_{nl}^I(w^2)=\sum_{k=0}^{N-1} \frac{w^{2k}}{k!}\frac{d^k}{dw^{2k}}
B_{nl}^I(0)+
\frac{w^{2N}}{\pi}\int_{4m_\pi^2}^\infty
ds \frac{\tan\delta_l^I(s)\,\mbox{Re}B_{nl}^I(s)}{s^N(s-w^2-i0)}\; .
\label{dr}
\ee
The solution of such type of dispersion relation was found long ago
by Omnes
\cite{omnes} and has the exponential form:

\be
B_{nl}^I(w^2)=B_{nl}^I(0)\exp
\biggl\{
\sum_{k=1}^{N-1} \frac{w^{2k}}{k!}\frac{d^k}{dw^{2k}}\log\, B_{nl}^I(0)
+
\frac{w^{2N}}{\pi}\int_{4m_\pi^2}^\infty
ds \frac{\delta_l^I(s)}{s^N(s-w^2-i0)}
\biggr\}\; .
\label{thesolution}
\ee
This equation, by derivation, is valid below the inelastic threshold
$w^2<16 m_\pi^2$, however the inelasticities (at least for
the isospin one
channel) in the
pion-pion scattering are small up to rather large energies.
Also the equation (\ref{thesolution}) being applied to the case of the pion
form factor ($n=0$, $l=1$) \cite{guerro,guerro1} gives
an excellent description
of experimental data up to $w^2\approx 2.5$~GeV$^2$
with two subtractions which are fixed by the value of the pion
e.m. form factor at zero  and its charge radius (see figure in
\cite{guerro1}). Moreover the solution (\ref{thesolution}) can be
systematically improved by taking into account higher thresholds
(see discussion in \cite{gunpb}), probably
the most important is the $K\bar K$ one.
The contribution of higher intermediate states can be suppressed
by choosing sufficiently large number of subtractions $N$, see
discussion in  section~\ref{reson}.

A great advantage of the solution eq.~(\ref{thesolution}) is that
it gives the $2\pi$DA's  in a wide range of energies in terms
of known $\pi\pi$ phase shifts and a few subtraction constants
(usually two is sufficient). The key observation is that these
constants are related to the {\it low-energy}
behaviour at $w^2\to 0$ of the $2\pi$DA's. In the low energy region
they can be computed using the effective chiral quark-pion lagrangian.
In the next section we shall compute the subtraction constants using
the low-energy effective lagrangian derived from the instanton model of QCD
vacuum \cite{DP86}. For the first estimates of the $2\pi$DA in this
model in low energy region see \cite{PW98}. After the subtraction
constants are fixed (see section~\ref{instantons}) we shall know the
$2\pi$DA's in a broad region of invariant masses.  Of particular
interest is the case when $w^2$ is close to the mass of some resonance
(say $\rho$): in this case we shall be able to extract DA of this
resonance from the solution (\ref{thesolution}),  this is
explained in the next subsection.

\subsection{Distribution amplitudes of resonances}
\label{reson}
The light-cone distribution amplitudes
of mesons are defined as matrix elements of
quark-antiquark non-local operators at light-like separations
\cite{ChZh84}. Here we consider as an example
the $\rho$-mesons DA's,
all equations in this section can be easily generalized to other
mesons.  Specifically, we consider chirally even and odd leading twist
$\rho$-meson DA's defined as:

\be
\langle
\rho^0(P)  | \, \bar\psi(x) \; \hat n \, \frac{\tau^3}{2} \; \psi(0)\,
|0\rangle \Bigr|_{x^+=x_\perp=0}=
(n\cdot e^{(\lambda)}) \frac{f_\rho^\| m_\rho}{ \sqrt 2}
\int_0^1 d z e^{-iz(P\cdot x)} \varphi_\rho^\|(z) ,
\label{dr1}
\ee
\be
\langle
\rho^0(P)  | \, \bar\psi(x) \; \sigma_{\mu\nu}n^\mu P^\nu
\, \frac{\tau^3}{2} \; \psi(0)\,
|0\rangle \Bigr|_{x^+=x_\perp=0}=
i(n\cdot e^{(\lambda)}) \frac{f_\rho^\perp m_\rho^2}{ \sqrt 2}
\int_0^1 d z e^{iz(P\cdot x)} \varphi_\rho^\perp (z) ,
\label{dr2}
\ee
where $e_\mu^{(\lambda)}$ is the polarization vector of the $\rho$-meson.
The normalization constants $f_\rho^\perp $ and $f_\rho^\| $
have dimension of mass and are chosen in such a way that
(below we suppress indices $\perp$ and $\|$ where it does
not lead to confusion):

$$
\int_0^1 dz \varphi_\rho(z)=1\; .
$$

The light-cone $\rho$ meson DA's are used to describe hard
leptoproduction of vector mesons \cite{mog}, and
in applications of light-cone QCD sum rules for exclusive semileptonic
and radiative weak decays of the $B$-meson \cite{Ali}.

There are certain problems with the  definitions
eqs.~(\ref{dr1},\ref{dr2}), because the $\rho$-meson
is an unstable particle.  One can treat the $\rho$-meson as a
quasistable particle, $i.e.$ consider the formal limit $\Gamma_\rho\to
0$.  From experimental point of view $\rho$-meson is detected as a
Breit-Wigner peak in the invariant mass distribution of produced pions.
Therefore a more adequate (and accurate) way to describe the
hard production experiments is in the terms of two-pion DA's.

In order to extract the rho meson DA's from the $2\pi$DA's we use the
expression (\ref{thesolution}) of $2\pi$DA in terms of the pion-pion
phase shifts. In the vicinity of the resonance the corresponding phase
shift (we shall speak for definiteness about $\rho$-meson, $i.e.$
$l=1$, isospin one) can be approximated by the Breit-Wigner formula:

\be
\delta_1^1(w^2\sim m_\rho^2)=\mbox{arctan}\bigl(
\frac{m_\rho \Gamma_\rho}{m_\rho^2-w^2}
\bigr)\; .
\ee
In the limit of vanishing width the contribution of
the resonance part
of the phase shift in the integral in eq.~(\ref{thesolution}) can be
written as :

\be
\lim_{\Gamma_\rho\to 0}
\frac{w^{2N}}{\pi}\int_{4m_\pi^2}^\infty
ds \frac{\delta_1^1(s)}{s^N(s-w^2-i0)}&=&
i\frac{\pi}{2} -\log(m_\rho^2-w^2)+
\log{m_\rho^2} \\
\nonumber
&+&({\rm pol.\ of \ order\ } N-1)\; .
\ee
This result being exponentiated gives rise to the meson pole.
Matching the
solution for  $2\pi$DA's eq.~(\ref{thesolution}) near $w^2\sim m_\rho^2$
to the Breit-Wigner form we get the
a relation between $\rho$ meson DA's and $2\pi$DA's. We obtain
the following expression for the coefficients of the expansion of
$\rho$ meson DA'S in Gegenbauer polynomials

$$
\varphi_\rho(z)=6 z(1-z)\biggl(
 1+\sum_{n={\rm even}}
a_n^{(\rho)} C_n^{3/2}(2 z-1) \biggr),
$$
in terms of subtraction constants entering eq.~(\ref{thesolution}) only

\be
a_n^{(\rho)}=B_{n1}(0)
\exp
\bigl(
\sum_{k=1}^{N-1} c_k^{(n1)} m_\rho^{2k}
\bigr)\; ,
\label{rho2pi}
\ee
where the subtraction constants $c_{k}^{(nl)}$ can be expressed
in terms of $B_{nl}(w^2)$ at low energies
\be
c_{k}^{(nl)}= \frac{1}{k!}\frac{d^k}{dw^{2k}}
[\log\, B_{nl}(w^2)-\log\, B_{l-1\,l}(w^2)]\Biggr|_{w^2=0}\;.
\label{ck}
\ee
The constants $f_\rho^{\|,\perp}$ can be computed as:

\be
\nonumber
f_\rho^\|=\frac{\sqrt 2 \Gamma_\rho {\rm Im}B_{01}^{\|}(m_\rho^2)
 }{g_{\rho\pi\pi}}\;,\\
f_\rho^\perp=\frac{\sqrt 2
\Gamma_\rho m_\rho
 {\rm Im}B_{01}^{\perp}(m_\rho^2) }{g_{\rho\pi\pi}f_{2\pi}^\perp}\;,
\label{dominance2}
\ee
where $g_{\rho\pi\pi}$ is a $\rho$ meson decay constant into two pions,
$f_{2\pi}^\perp$ is defined by eq.~(\ref{f2pi} ).
The constants $B_{01}^{\|,\perp}(m_\rho^2)$ are purely imaginary and
can be computed using eq.~(\ref{thesolution}) in terms of the pion-pion
phase shift $\delta_1^1$ and low-energy subtraction constants. The
latter are actually pion charge radius and radius of pion tensor form
factor (they are estimated in the next section).

eqs.~(\ref{rho2pi},\ref{dominance2}) can be easily generalized to
the DA's of higher spin resonances. For example, the chirally even
DA's of an
isovector resonance of spin $j$ can be computed in terms of
the $2\pi$DA's as:

\be
\phi_j(z)=6 z(1-z)\biggl( C_{j-1}^{3/2}(2 z-1) +
\sum_{n=j+1}^\infty a_n^{(j)} C_{n}^{3/2}(2 z-1)
\biggr)\; ,
\ee
with coefficients $a_n^{(j)}$ computed as:

\be
a_n^{(j)}=\frac{B_{nj}(0)}{B_{j-1\,j}(0)}
\exp
\bigl(
\sum_{k=1}^{N-1} c_k^{(nj)} m_R^{2k}
\bigr)\; .
\label{anyspin}
\ee
The corresponding normalization constant is:

\be
f_R
=\frac{\sqrt 2 \Gamma_R {\rm Im}B_{j-1\, j}(m_R^2)
 }{g_{R\pi\pi}}\;.
\label{anyspinf}
\ee
The scale dependence of $f_R$ and $a_n^{(j)}$ is obvious.

We see that the $\rho$ meson DA's (we speak for definiteness about
$\rho$ meson, but all arguments below apply also to any
other resonance)
are fixed in terms of the subtraction
(low-energy) constants, which would be all fixed if we knew the exact
low-enregy lagrangian of QCD\footnote{For example, the
constant $c_{1}^{\|(01)}$ is proportional to the parameter $L_9$ of
Gasser-Leutwyler parametrization of the EChL \cite{GL}.}. The latter is
not fully known, so we shall resort to the low-energy
lagrangian derived from the instanton model of the QCD vacuum
\cite{DP86}.

The r.h.s. of eq.~(\ref{rho2pi}) depends on the number of subtractions
$N$ whereas its l.h.s. is independent of this number.
At first glance
this fact implies that the subtraction constants $c_k^{(nl)}$ are all
zero. However we should keep in mind that the dispersion relation
eq.~(\ref{dr}) is not exact, in its derivation the contribution of
intermediate states other than two pion was neglected. In order to
suppress the contribution of the higher states we have to take a
large enough number of subtractions in the dispersion relation
(\ref{dr}).
This implies that one can expect an (approximate) independence of the r.h.s.
 of eq.~(\ref{rho2pi}) of $N$ only for sufficiently large $N$.

Increasing the number of
subtractions suppresses the high energy
region, where the approximations were made, and hence increases the
accuracy of the approximate solution (\ref{thesolution}). However,
increasing the number of subtractions we emphasize the low energy
region the quantitative description  of which is limited by incomplete
knowledge of the effective chiral lagrangian. Therefore for
phenomenological applications one has to choose some optimal $N$.
The analysis
of the pion e.m. form factor shows \cite{guerro,guerro1} that already
for $N=2$ the contribution of the high energy states
is sufficiently suppressed to describe the pion form factor with good
accuracy up to $w^2\approx 2.5$~GeV$^2$.

In the next section we
take $N=2$. Since with this value the
contribution of the high energy states in dispersion relation
(\ref{dr}) is sufficiently well damped and simultaneously we can
reliably use the approximate low energy quark-pion chiral lagrangian
derived from the instanton model of QCD vacuum \cite{DP86}. In language
of the chiral perturbation theory $N=2$ corresponds to the fixing of
the effective chiral lagrangian (EChL) to the fourth order. To this
order the effective chiral theory derived from the instanton model
gives \cite{DP86,DPP} the constants of the fourth order EChL (actually
only its part describing pions interacting with external vector and
axial currents) which are very close to ones obtained in the
phenomenological analysis of Gasser and Leutwyler \cite{GL}.

\section{Fixing low-energy constants from instantons}
\setcounter{equation}{0}
\label{instantons}

The technique how to compute matrix elements of bilocal
quark operators at low energies
in the effective low-energy quark-pion theory was
explained in details in refs.~\cite{PP,PPRWG98} (for pion and photon
DA's) and in \cite{PW98} (for $2\pi$ DA's). Here we only sketch the
main points of the calculations, for details the reader should consult
refs.~\cite{PP,PPRWG98,PW98}.

We shall compute the
$2\pi$DA's at small $w^2$,
eq.~(\ref{definition1},\ref{definition2}), using the effective
low-energy
theory of pions interacting with massive ``constituent'' quarks, which
has been derived from the instanton model of the QCD vacuum
\cite{DP86}. The coupling of the pion field to the quarks, whose form is
restricted by the chiral invariance, is described by the action
\be
S_{\rm int}&=& \int d^4 x\; \bar \psi(x) \sqrt{M(\partial^2)}
\; U^{\gamma_5} (x) \;
\sqrt{M(\partial^2)}\psi(x) ,
\label{action}
\ee
where
\be
U^{\gamma_5} (x) &=& e^{i \gamma_5 \tau^a \pi^a (x) / f_\pi}
\;\; = \;\; 1 \; + \; \frac{i}{f_\pi}\gamma_5\pi^a(x)\tau^a-
\frac{1}{2f_\pi^2}\pi^a(x)\pi^a(x) \;\; + \;\; \ldots .
\ee
Here, $f_\pi = 93 \, {\rm MeV}$ is the weak pion decay constant

The momentum--dependent dynamical quark mass, $M(p)$, plays the role of
an UV regulator. Its form for Euclidean momenta was derived in
ref.\cite{DP86}.  The momentum dependent mass cuts loop integrals at
momenta of order of the inverse average instanton size,
$\bar\rho^{-1} \approx 600$~MeV.  One should note that the value of the
mass at zero momentum, $M(0)$, is parametrically small compared to
$\bar\rho^{-1}$; the product $M(0) \bar\rho$ is proportional to the
packing fraction of the instantons in the medium, which is a small
algebraic parameter fundamental to this picture.  Numerically, a value
$M(0) = 350 \, {\rm MeV}$ was obtained in ref.~\cite{DP86}.

\subsection{  Isovector $2\pi$DA's}
Using the technique described in \cite{PW98} we can easily see
that the loop integral in the effective theory for
the chirally odd
isovector $2\pi$DA (\ref{definition2}) contains no divergencies, and
hence we can neglect during its computation the momentum dependence of
the constituent quark mass. Chirally even isovector $2\pi$DA
(\ref{definition1}) was computed in \cite{PW98}, the computation there
was done neglecting the momentum dependence of
the constituent quark mass
and absorbing the divergencies into $f_\pi$. It is easy to see that
all (logarithmic) divergencies are contained in the $w^2$-independent
piece of this DA, the $w^2$ dependent piece is finite.  The result of
\cite{PW98} in this approximation for $w^2=0$ was very simple:

\be
\Phi_{\|} (z, \zeta, w^2=0 )=\mbox{sign}(1-2 \zeta)
\theta[(z-\zeta) (1-\zeta-z)]\;,
\ee
where $\theta(x)$ is a step function. We see that chirally even
isovector $2\pi$DA exhibits jumps at points $z=\zeta$ and $z=1-\zeta$.
As was shown in \cite{PPPBWG,PP,PPRWG98}
the momentum dependence of the constituent quark mass becomes crucial at
these points. Here
for the numerical estimates of the $w^2$ independent piece
we  employ a simple numerical fit
to the momentum dependent constituent quark mass obtained from the
instanton model of the QCD vacuum,
\be
M(-p^2) &=& \frac{M_0}{\left( 1 + 0.5 \, p^2 \bar\rho^2 \right)^2}.
\ee
As we already mentioned the $w^2$-dependent piece of the chirally even
isovector $2\pi$DA is finite (it does not exhibit the jumps) and hence
it can be computed neglecting the UV-cutoff provided by the momentum
dependence of the constituent quark mass.

\begin{figure}
\setlength{\epsfxsize}{15cm}
\setlength{\epsfysize}{15cm}
\epsffile{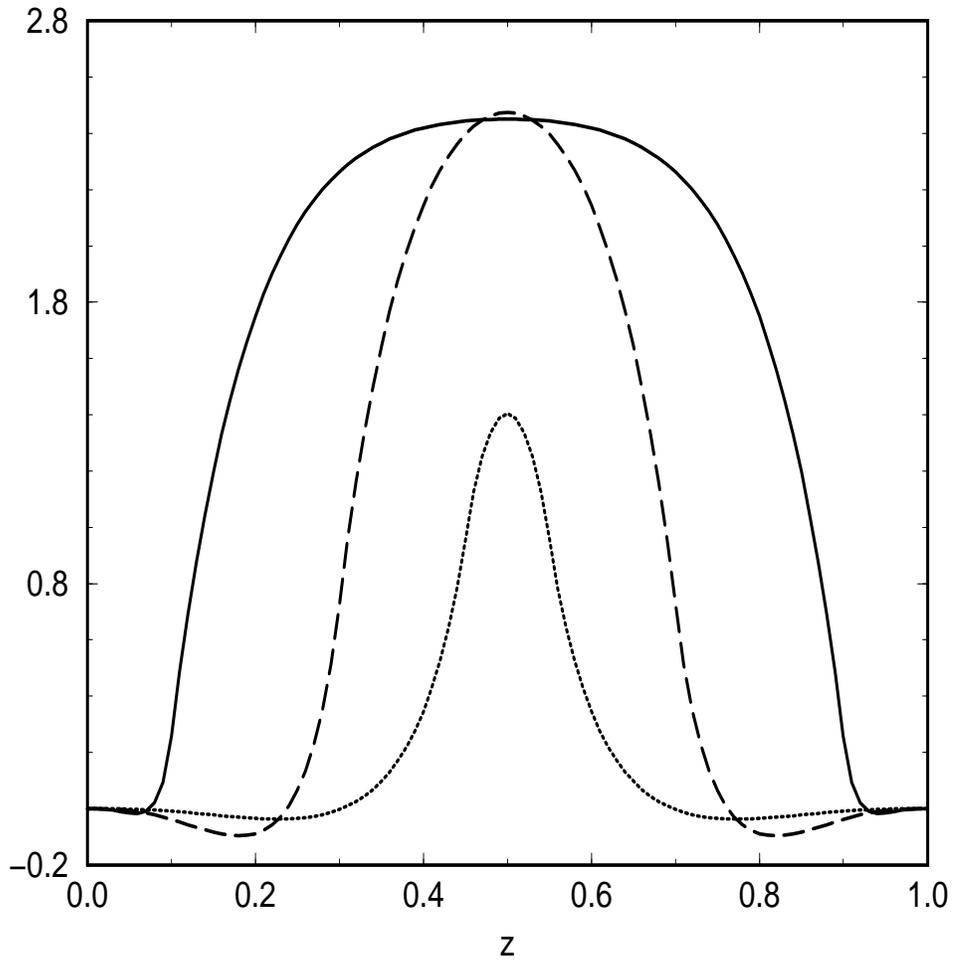}
\caption[]
{The two--pion chirally even isovector distribution amplitude
$\Phi_\|^{I=1}(z,\zeta,w^2=0)$ at various values of $\zeta$.
{\it Solid line:} $\zeta=0.1$.
{\it Dashed line:} $\zeta=0.3$.
{\it Dotted line:} $\zeta=0.45$}
\label{fig_fig2}
\end{figure}

Results of a computation of the
chirally even isovector $2\pi$DA at $w^2=0$ are shown in
Fig.~\ref{fig_fig2}, where we have plotted
$\Phi_{\|}^{I=1} (z, \zeta, w^2=0 )$ as a function of $z$ at various
values of $\zeta$. We see that the function is concentrated in the
interval of $z$ between $\zeta$ and $1-\zeta$ and the
jumps are
smoothed. Also it is interesting to note that if we take $\zeta=1$
the result for $\Phi_{\|}^{I=1}(z, \zeta, w^2=0 )$ coincides exactly with
that for the pion DA obtained in \cite{PP,PPRWG98} in the same model,
what is not surprising due to the soft pion theorem (\ref{letwf}).

Inspecting loop integrals for the  chirally even
$2\pi$DA's in the model
(see typical expressions in ref.~\cite{PW98}) we can make
interesting observation: in the limits $\zeta\to 1$ and $m_\pi\to 0$
the $w^2$ dependence appear exclusively in
the combination $z(1-z)\,w^2$.
This observation allows us to obtain the following model relations
for the coefficients of Gegenbauer expansion (\ref{razhl})
$B_{nl}^{\|}(w^2)$ (we omit the superscript $\|$ below
to make the expression not so busy):

\be
\sum_{
\scriptstyle {l=1} \atop {\rm odd } }^{n+1} \frac{d^j}{dw^{2j}}
B_{nl}^{I=1}(w^2)\bigr|_{w^2=0}&=&0\;,
\quad  {\rm for\ any \ even \ } n\geq 2j \; ,\\
\sum_{
\scriptstyle {l=0} \atop {\rm even } }^{n+1} \frac{d^j}{dw^{2j}}
B_{nl}^{I=0}(w^2)\bigr|_{w^2=0}&=&0\;,
\quad  {\rm for\ any \ odd \ } n\geq 2j-1 \; .
\ee
These $model$ results, in a sense, can be viewed as a generalization
of the soft pion theorems (\ref{letcoef}, \ref{let0}).
Although we find  these relations in the model calculations, we can
speculate that they have more
general nature and follow from the chiral
Ward identities in the limit of large colour number $N_c\to\infty$.
This conjecture will be worked out in details elsewhere \cite{prep}.

Below we give the results of numerical calculations of a
few first coefficients
of Gegenbauer expansion (\ref{razhl}) $B_{nl}^\|(w^2)$ of the isovector
chirally even $2\pi$DA at low $w^2$:

\be
\label{emff}
B_{01}^\|(w^2)&=&1+\frac{N_c\,w^2}{24\pi^2f_\pi^2}+\ldots\; ,\\
B_{21}^\|(w^2)&=&-0.2-\frac{7N_c\,w^2}{1440\pi^2f_\pi^2}+\ldots\; ,\\
B_{23}^\|(w^2)&=&0.26+\frac{7N_c\,w^2}{1440\pi^2f_\pi^2}+\ldots\; ,\\
B_{41}^\|(w^2)&=&-0.006-\frac{11N_c\,w^2}{25200\pi^2f_\pi^2}+
\ldots\; ,\\
B_{43}^\|(w^2)&=&-0.12-\frac{11N_c\,w^2}{10800\pi^2f_\pi^2}+\ldots\; ,\\
\label{b45}
B_{45}^\|(w^2)&=&0.14+\frac{11N_c\,w^2}{7560\pi^2f_\pi^2}+\ldots\; .
\ee
From this calculation we immediately can extract the coefficients of
the Gegenbauer expansion of the pion DA (\ref{razhlpi})
using the soft pion theorem (\ref{letcoef}):

\be
a_2^{(\pi)}=0.06,\quad a_4^{(\pi)}=0.014 \;.
\label{a2a4num}
\ee
These are exactly the values obtained in \cite{PPRWG98}.

Remembering that $B_{01}^\|(w^2)=F_\pi(w^2)$ we extract
from eq.~(\ref{emff}) the value of the
pion electromagnetic charge radius
(without chiral loops!)
\be
\langle r^2\rangle_{\rm e.m.}=
\frac{N_c}{4\pi^2 f_\pi^2} ,
\label{chr}
\ee
which coincides with the result obtained previously in ref.~\cite{DP86}.
The model result (\ref{chr}) numerically gives
$\langle r^2\rangle_{\rm e.m.}=0.35$~fm$^2$ to be compared with the
experimental value $0.439\pm 0.008$~fm$^2$ \cite{expchr}.
Comparing these numbers we
should keep in mind that the model calculations do not include the
chiral logarithms, the latter appear when we perform the integral in
eq.~(\ref{thesolution}) (for $n=0$, $l=1$) with the pion-pion phase
shift obtained in the leading order of the chiral perturbation theory:

$$
\delta_1^1(w^2)= \frac{w^2 (1-4 m_\pi^2/w^2)^{3/2}}{96 \pi
f_\pi^2} \;.
$$
Then the result for the pion charge radius becomes (remember that
$B_{01}(w^2)=F_{\pi}(w^2 )$):

$$
\langle r^2\rangle_{\rm e.m.}=
\frac{N_c}{4\pi^2 f_\pi^2}-\frac{1}{16\pi^2 f_\pi^2}\log
(\frac{m_\pi^2}{m_\rho^2}) \approx 0.45\ {\rm fm}^2\; ,
$$
in good
agreement with the experimental value.

Basing on the results (\ref{emff}-\ref{b45}) we can easily
estimate the low-energy subtraction constants, see
eqs.~(\ref{thesolution},\ref{ck}),
the results are summarized in Table~1, they refer to
the normalization point $\mu\sim 1/\bar \rho\approx 600$~MeV the
inverse instanton size in the instanton vacuum.

Having the values of the low-energy subtraction constants we can
obtain the chirally even isovector $2\pi$DA in a wide interval of
invariant two-pion masses using the solution (\ref{thesolution})
with $N=2$ (see the discussion in previous section).
The first qualitative observation we can make on basis of Table~1 is
that the coefficients in front of $C_{n}^{3/2}(2 z-1)$ for $n=2,4$
(they describe the deviation of DA's from its asymptotic form)
for the $P$-wave production of pions are negative and $B_{41}^\|(w^2)$
increases (by absolute value) whereas $B_{21}^\|(w^2)$ decreases
with increasing of $w^2$.

Here we quote the
result for the chirally even $\rho$ meson DA extracted from $2\pi$DA
using eqs.~(\ref{rho2pi},\ref{dominance2}):

\be
\varphi_\rho^\|(z)=6 z(1-z)\bigl[1-0.14\, C_2^{3/2}(2 z-1)-
0.01\, C_4^{3/2}(2 z-1)+\ldots \bigr]\;,
\label{ourrhowf}
\ee
with normalization constant $f_\rho^\|=190$~MeV according to
eq.~(\ref{dominance2}) in a good agreement with
experimental value $f_\rho^\|=195\pm 7$~MeV \cite{PDG}.
The $\rho$ meson DA's were the subject
of the QCD sum rules calculations \cite{ChZh84,BB,BM}, our result
eq.~(\ref{ourrhowf}) is in a qualitative disagreement with the
results of QCD sum rule
calculations, the sign of $a_2^\|$ obtained here
is opposite to the QCD sum rules
predictions $a_2^\|=0.18\pm 0.1$ \cite{BB} and $a_2^\|=0.08
\pm 0.02$, $a_4^\|=-0.08\pm 0.03 $ \cite{BM} (these results refer
to normalization point $\mu=1$~GeV).
The difference in the sign of ours and QCD sum rule calculations
can be traced back to the signicicantly different values
of $a_2^{(\pi)}$. Indeed, using the eq.~(\ref{rho2pi}) and the crossing
relation (\ref{therelation}) we can relate $a_2^{(\rho)}$,
$a_2^{(\pi)}$ and the third moment of the quark distribution in a pion
in the following way:

\be
\nonumber
a_2^{(\rho)}=B^{I=1}_{21}(0)\exp(c_1^{(21)}m_\rho^2)&=&
(a_2^{(\pi)}-B^{I=1}_{23}(0))\exp(c_1^{(21)}m_\rho^2)\\
&=&
(a_2^{(\pi)}-\frac 76 M_3^{(\pi)})\exp(c_1^{(21)}m_\rho^2)\, .
\ee
The value of $a_2^{(\pi)}$ in the instanton model of the QCD vacuum
is very small (see eq.~(\ref{a2a4num}) and ref.~\cite{PPRWG98}), such
that $a_2^{(\pi)}<\frac 76 M_3^{(\pi)}$. On contrary the value
of $a_2^{(\pi)}$ in the QCD sum rule calculations \cite{BraunFilyanov,
RadyushkinMikhailov} larger than $\frac 76 M_3^{(\pi)}$, what implies
the positive sign of $a_2^{(\rho)}$.

Using eq.~(\ref{rho2pi})
and the value of the subtraction constants from Table~1 we can make a
rough estimate of the shape of the
$\rho'$ chirally even DA ($m_{\rho'}=1.465$~GeV):

\be
\varphi_{\rho'}^\|(z)=6 z(1-z)\bigl[1-0.055\, C_2^{3/2}(2 z-1)-
0.064\, C_4^{3/2}(2 z-1)+\ldots \bigr]\;.
\ee
This is just a rough estimate because the mass of $\rho'$ is rather
big and importance of neglected in the
derivation of eq.~(\ref{thesolution})
high energy states increases and hence the accuracy of the solution
eq.~(\ref{thesolution}) becomes poor. But even from this rough
estimate we see an interesting tendency: the coefficients
in front of the Gegenbauer polynomials with
$n>2$ in the expansion of $\rho'$ meson
chirally even DA become comparable with the leading
nontrivial
($n=2$)
coefficient and the DA develops  nodes.

Now we turn to results for the chirally odd $2\pi$DA
eq.~(\ref{definition2}). Simple calculations of matrix element
eq.~(\ref{f2pi}) give the following value of the constant $f_{2\pi}$

\be
f_{2\pi}=\frac{8\pi^2 f_\pi^2}{N_c M}\approx 650\ {\rm MeV}\;.
\ee

Calculations of $B_{nl}^\perp(w^2)$ in the low energy region are
straightforward (details will be published elsewhere \cite{prep}),
the results are:

\be
\label{ttff}
B_{01}^\perp(w^2)&=&1+\frac{w^2}{12 M^2}+\ldots\; ,\\
B_{21}^\perp(w^2)&=&\frac{7}{36}\bigl(1
-\frac{w^2}{30 M^2}+\ldots\;\bigr) ,\\
B_{23}^\perp(w^2)&=&\frac{7}{36}\bigl(1
+\frac{w^2}{30 M^2}+\ldots\;\bigr) ,\\
B_{41}^\perp(w^2)&=&\frac{11}{225}\bigl(1
-\frac{5w^2}{168 M^2}+\ldots\;\bigr) ,\\
B_{43}^\perp(w^2)&=&\frac{77}{675}\bigl(1
-\frac{w^2}{630 M^2}+\ldots\;\bigr) ,\\
B_{45}^\perp(w^2)&=&\frac{11}{135}\bigl(1
+\frac{w^2}{56 M^2}+\ldots\;\bigr) \;.
\ee
From these equations we get immediately the values of the subtraction
constants $c_{k}^{\perp (nl)}$, they are summarized in Table~1. The
value of ``tensor'' isovector radius of pion is from eq.~(\ref{ttff})
(without chiral loops!):

\be
\langle r^2 \rangle_t= \frac{1}{2 M^2} \approx 0.16\ {\rm fm}^2\; .
\ee
The chiral loops are the same as for charge radius.

Using the results from Table~1 for $c_{k}^{\perp (nl)}$ and
eqs.~(\ref{rho2pi},\ref{dominance2}) we obtain the chirally odd
$\rho$-meson DA:

\be
\varphi_{\rho}^\perp(z)=6 z(1-z)\bigl[1+0.11\, C_2^{3/2}(2 z-1)
+0.03\, C_4^{3/2}(2 z-1)+\ldots \bigr]\;,
\ee
and the normalization constant $f_\rho^\perp$:

\be
f_\rho^\perp=f_\rho^\|\,  \frac{m_\rho}{f_{2\pi}}\exp[
(\langle r^2 \rangle_t-\langle r^2\rangle_{\rm e.m.})
m_\rho^2/6]\approx
140\ {\rm MeV}\;.
\ee
[This value refers to the normalization point
$\mu\sim 1/\bar \rho\approx 600$~MeV]. The
QCD sum rule calculations \cite{BB,BM} give $f_\rho^\perp=160\pm 10$~MeV
$a_2^\perp=0.2\pm 0.1$ \cite{BB}
and $f_\rho^\perp=169\pm 5$~MeV
$a_2^\perp=0.36\pm 0.03$  \cite{BM}
at the normalization point $\mu=1$~GeV
in fair agreement with our results.

The results
for parameters of the $\rho$ meson DA's are summarized in Table~2
where we also quote the
results of the QCD sum rule calculations \cite{BB,BM}.
With the help of Table~1 and eqs.~(\ref{thesolution},\ref{anyspin})
the reader can easily construct DA's of isovector mesons of any spin.

\subsection{ Note on isoscalar $2\pi$DA's}
The quark isoscalar chirally even
$2\pi$DA's under ERBL evolution \cite{ERBL}
mix with gluon $2\pi$DA's. In the instanton vacuum model
considered here the gluon $2\pi$DA's are parametrically small
at low normalization point ($\mu\sim 1/\bar \rho$)
in the instanton packing fraction (see ref.~\cite{DPW96})
and hence are zero at the level of accuracy considered here.

The second moment of the chirally even $2\pi$DA eq.~(\ref{definition1})
is related to the quark part of the (symmetric) energy-momentum tensor
$T_q^{\mu\nu}$:

\be
\int_0^1 dz (2\,z -1)\, \Phi_\|^{I=0}(z,\zeta,w^2) =
\frac{1}{2 P^{+2}} {}_{out} \langle \pi^a(p_1) \pi^b(p_2)|
T_q^{++}|0\rangle \; .
\label{m2}
\ee
In the instanton model of the QCD vacuum (see \cite{DPW96})
the gluon part  of the energy-momentum tensor (projected on the
light-cone direction $n^\mu$ !) is parametrically small in the packing
fraction of the instanton liquid. Therefore here we can use, the
so-called, quark-antiquark approximation, $i.e.$
$T_q^{++}=T^{++}+$~small corrections. At this level of accuracy we
have ({\it cf.} equation (\ref{norm1})):

\be
\int_0^1 dz (2\,z -1)\, \Phi_\|^{I=0}(z,\zeta,w^2=0) =
-2 \zeta(1-\zeta) \;.
\ee

The model calculations of the isoscalar chirally even $2\pi$DA
give the following result for its second moment:

\be
\int_0^1 dz (2\,z -1)\, \Phi_\|^{I=0}(z,\zeta,w^2) =
-2 \zeta(1-\zeta)
\bigl[1+\frac{N_c w^2}{48\pi^2 f_\pi^2}+\ldots \bigr] \;.
\label{m2m}
\ee
We see that the overall normalization is fixed by the energy momentum
conservation and the radius of pion form factor of the energy momentum
tensor is:

\be \langle r^2\rangle_{{\rm EMT}}=\frac{N_c }{8\pi^2
f_\pi^2}= 0.18\,{\rm fm}^2\;.
\ee

From eq.~(\ref{m2m}) we get immediately the values of the low-energy
subtraction constant:

\be
B_{10}^\|(0)=-\frac{5}{9}, \quad B_{12}^\|(0)=\frac{5}{9}\;,
\ee
\be
\frac{d}{d\,w^2}\log B_{10}^\|(w^2)\bigl|_{w^2=0}=
\frac{d}{d\,w^2} \log B_{12}^\|(w^2)\bigl|_{w^2=0}=
\frac{N_c }{48\pi^2 f_\pi^2}=0.73\;{\rm GeV}^{-2}\;.
\label{mod0}
\ee
Obviously the soft pion theorem
eq.~(\ref{let0}) is satisfied.
Let us use the above model calculations and crossing relation
(\ref{relation}) to estimate the low momentum transfer behaviour of
the second moments of skewed quark distribution in a pion.
From eq.~(\ref{relation}) for the second moment we have
\be
\nonumber
\int_{-1}^1 dX X H^{I=0}(X,\xi,t)&=&
\frac 3 5 B^\|_{10}(t)\ \xi^2 +
\frac 3 5
B^\|_{12}(t)\ \xi^2\ C_2^{1/2}\biggl(\frac{1}{\xi}\biggr)\\
 &=&\frac 3
5\bigl[ B^\|_{10}(t)\ \xi^2+B^\|_{12}(t)\frac{3-\xi^2}{2} \bigr]\;
.  \ee Now if we substitute in the above equation the result of the
model calculation (\ref{mod0}) we get:

\be
\int_{-1}^1 dX X H^{I=0}(X,\xi,t)=
\frac 12 (1-\xi^2)
\bigl[1+\frac{N_c t}{48\pi^2 f_\pi^2}+\ldots \bigr]\; .
\ee

\section{Two-pion distribution amplitudes in hard pions production
processes}
\setcounter{equation}{0}
\label{hard}

The dependence of the two-pion hard production amplitude
(\ref{proc2}) on the $2\pi$DA's factorizes into the factor:
\be
A\propto \int_0^1\frac{dz}{z(1-z)} \,\Phi^{I=1}_\|(z,\zeta,w^2; \bar
Q^2)\;, \ee
for the two pions in the isovector state, and

\be
A\propto \int_0^1 dz\frac{(2 z-1)}{z(1-z)}\,
\Phi^{I=0}_\|(z,\zeta,w^2; \bar Q^2)\;,
\ee
for the pions in the isoscalar state ($e.g.$ for the
$\pi^0\pi^0$ production).  In the above equations we showed also
the dependence of the $2\pi$DA's on the scale of the process $\bar Q^2$,
which is governed by the ERBL evolution equation \cite{ERBL}.

For the process (\ref{proc2}) at small $x_{Bj}$
(see $e.g.$ recent measurements \cite{zeus}) the production of two
pions in the isoscalar channel is strongly suppressed
relative to the isovector channel because the former is
mediated by $C$-parity odd exchange. At asymptotically large $Q^2$
one can use the asymptotic form of isovector $2\pi$DA:

\be
\lim_{\bar Q^2\to\infty}\Phi^{I=1}_\|(z,\zeta,w^2; \bar Q^2)=
6 F_\pi(w^2) z(1-z) (2\zeta-1)\; ,
\ee
where $F_\pi(w^2)$ is the pion e.m. form factor in the time-like region.
Therefore the shape of
$\pi^+\pi^-$ mass spectrum in the hard electroproduction process
(\ref{proc2}) at small $x_{Bj}$ and asymptotically large $Q^2$ should
be determined completely by the
pion e.m. form factor in time-like region:

\be
\lim_{Q^2\to\infty} A \propto e^{i\delta_1^1(w^2)}|F_\pi(w^2)| (2
\zeta-1) \;.
\label{asymf}
\ee
Deviation of the $\pi^+\pi^-$ mass spectrum from its asymptotic form
eq.~(\ref{asymf})
(``skewing") can be parametrized at small $x_{Bj}$
and large $\bar Q^2$ in the form:

\be
\nonumber
A\propto e^{i\delta_1^1(w^2)}|F_\pi(w^2)|\Bigl[
1+B_{21}(0;\mu_0)\exp\{c_1^{(21)} w^2\}
\biggl(
\frac{\alpha_s(\bar Q^2)}{\alpha_s(\mu_0)}
\biggr)^{50/(99-6 N_f)}
\Bigr] (2\zeta-1)+\\
\nonumber
e^{i\delta_3^1(w^2)}B_{23}(0;\mu_0)\exp\{b_{23} w^2+R_3^1(w^2)\}
\biggl(
\frac{\alpha_s(\bar Q^2)}{\alpha_s(\mu_0)}
\biggr)^{50/(99-6 N_f)}
C_3^{1/2}(2\zeta-1)\\
+\mbox{\rm higher\ powers\ of\ }1/\log(\bar
Q^2) \;,
\ee
here
\be
R_l^I(w^2)= \mbox{\rm Re}
\frac{w^{4}}{\pi}\int_{4m_\pi^2}^\infty
ds \frac{\delta_l^I(s)}{s^2(s-w^2-i0)}\ \ {\rm and\ } \
b_{nl}=
\frac{d}{dw^2}
\log\, B_{nl}(w^2)\Biggr|_{w^2=0}\;.
\label{paramf}
\ee
We see that the deviation of the $\pi\pi$ invariant mass spectrum from
its asymptotic form eq.~(\ref{asymf}) in this approximation is
characterized by a few low-energy constants ($B_{21}(0)$, $B_{23}(0)$,
$c_1^{(21)}$, $b_{23}$), other quantities-- the pion e.m. form factor
and the $\pi\pi$ phase shifts--are known from low-energy experiments.
In principle, using the parametrization (\ref{paramf}) one can extract
the values of these low-energy parameters from the shape of $\pi\pi$
spectrum (not absolute cross section!) in diffractive production
experiments. Knowing them one can obtain the deviation of the $\pi$
meson DA (see eq.~(\ref{letcoef}))
\be \nonumber a_2^{(\pi)}=B_{21}(0)
+ B_{23}(0) \;, \ee
and the $\rho$ meson DA (see eq.~(\ref{rho2pi}))
\be
\nonumber
a_2^{(\rho)}=B_{21}(0) \exp(c_1^{(21)}m_\rho^2)  \;,
\ee
from their asymptotic form $6z(1-z)$, as well as the normalization
constants for the DA of isovector resonances of spin three (see
eq.~(\ref{anyspinf})). Additionally the parameter $B_{23}(0)$ is
related by the crossing relations to the third moment of the valence
quark distribution in a pion (see eq.~(\ref{f2tp})):
\be
M_3^{\pi}=\frac{6}{7} B_{23}(0)\, .
\ee
We see that, in principle, the diffractive production of two pions can
be used to get an information about quark distribution in a pion.

In analysis of experiments on two pion diffractive production
off nucleon (see e.g. \cite{zeus}) the non-resonant background
is described by S\"oding parametrization \cite{soding}, which
takes into account rescattering of produced pions on final
nucleon. Let us note however that in a case of hard ($Q^2\to \infty$)
diffractive production the final state interaction of pions with
residual nucleon is suppressed by powers of $1/Q^2$ relative to the
leading twist amplitude.
Here we proposed alternative leading-twist parametrization
(\ref{paramf}) describing the so-called ``skewing" of two pion
spectrum.

\section{Conclusions}
\setcounter{equation}{0}
\label{sec_conclusions}
We showed that the
two-pion distribution amplitudes are the most general object
to describe the
hard electroproduction of two pions. Using Watson
final state interaction theorem and
the soft pion theorems proved here, we
can determine the $2\pi$DA's in a wide region of pion invariant masses
in term of the pion phase shifts and a few low-energy (subtraction)
constants.  The former are know from data on $\pi\pi$ scattering (soft
physics), the latter are non-perturbative input which we estimated here
using the instanton model of the QCD vacuum. We showed that these
non-perturbative characteristic can, in principle, be extracted from
the shape of $\pi\pi$ mass spectra in
diffractive pions production experiments. Then they can be used to
determine, say, the deviation of the $\pi$ meson DA (see
eq.~(\ref{letcoef})) and $\rho$ meson DA (see eq.~(\ref{rho2pi})) from
their asymptotic form $6z(1-z)$, and hence to obtain non-perturbative
information about structure of mesons.  We demonstrated (see
section~\ref{reson}) that DA's of resonances of any spin can be
determined from $2\pi$ DA's.

We derive the crossing relation which relate various distribution
amplitudes to the quark distributions in a pion.

The soft pion theorems considered in section~\ref{softpi} give an example
of how low-energy (chiral) physics can be studied in hard processes
(see recent work in this direction \cite{EFS}).

The methods developed here can be easily generalized for more
complicated cases. One of examples is generalized $N\to \pi N$
skewed parton distributions \cite{FPS,MVPB98} entering the QCD
description of the ``non-diagonal" deeply--virtual Compton scattering
$\gamma^\ast_L +p \to \gamma + \pi N$.

\section{Acknowledgments}
I am grateful to P. Ball, V.Braun, L.~Frankfurt, K.~Goeke,
P.~Pobylitsa, A.~Radyushkin,  A.~Sch\"afer, M.~Strikman, O.~Teryaev and
C.~Weiss for inspiring discussions.
Special thanks to D.I. Diakonov for many valuable remarks.
This work has been supported in
part by the BMFB (Bonn), Deutsche Forschungsgemeinschaft (Bonn) and by
COSY (J\"ulich).

\newpage

\appendix
\renewcommand{\theequation}{\Alph{section}.\arabic{equation}}

\section{Appendix}
\setcounter{equation}{0}

Due to the kinematical constraint (\ref{con}) the physical region
of variable $\zeta$ is bounded to the interval \\
$\zeta_- \leq \zeta\leq\zeta_+$, where

\be
\zeta_\pm=\frac 12\biggl(
1 \pm \sqrt{1-\frac{4 m_\pi^2}{w^2}}
\biggr) \; .
\ee
It implies that the Gegenbauer polynomials $C^{1/2}_l(2\zeta-1)$
used in the expansion (\ref{razhl}) are not orthogonal on the interval
$\zeta_- \leq \zeta\leq\zeta_+$. The correct set of polynomials in this
case is:
\be
{\cal P}_l(\zeta)=C^{1/2}_l\bigl(\frac{2\zeta-1}{\sqrt{1-\frac{4
m_\pi^2}{w^2}}}\bigr).
\label{newset}
\ee
It is easy to check that these polynomials are orthogonal on
the interval $\zeta_- \leq \zeta\leq\zeta_+$:

\be
\int_{\zeta_-}^{\zeta_+} d \zeta\ {\cal P}_l(\zeta) {\cal P}_k(\zeta)
=\left\{
\begin{array}{cc}
0\, ,& \hspace{0.5cm}  l\neq k \,, \\
\frac{1}{2l+1}\ \sqrt{1-\frac{4 m_\pi^2}{w^2}}
,& \hspace{0.5cm} l=k \,.
\end{array}
\right.
\ee
In the case of $w^2$ close to the threshold $4 m_\pi^2$ the expansion
(\ref{razhl}) is modified:

\be
\Phi_{\|,\perp} (z, \zeta, w^2 )=6z(1-z)\sum_{n=0}^{\infty}
\sum_{l=0}^{n+1} B_{nl}^{\|,\perp}(w^2) C_n^{3/2}(2 z-1)
{\cal P}_l(\zeta).  \label{razhl1} \ee
From the normalization condition (\ref{norm}) we conclude that

\be
B_{01}^{\|}= \sqrt{1-\frac{4 m_\pi^2}{w^2}} F_\pi^{\rm e.m.}(w^2)\, .
\ee

Also the equations in section~\ref{tri} are modified near threshold.
The phase space volume is:

\be
\int d({\rm phase\ space}) =\frac{1}{64\pi^2}
\int_{\zeta_-}^{\zeta_+} d\zeta'
\int_0^{2\pi} d\theta_\perp\, .
\ee
The scattring angle in the c.m. system is expressed in terms
light cone variables as following

\be
\nonumber
\cos(\theta_{cm})&=&\frac{1}{1-4 m_\pi^2/w^2}
\Biggl[
1-2(\zeta+\zeta'-2\zeta\zeta')\\
&-&
4\sqrt{(\zeta(1-\zeta)-m_\pi^2/w^2)(\zeta'(1-\zeta')-m_\pi^2/w^2)}
\cos(\theta_\perp)
\Biggr]\; .
\label{cmlc1}
\ee
With this equality the expression (\ref{additionth}) is modified as
follows:

\be
\int_0^{2\pi} d\theta_\perp P_l[\cos(\theta_{cm})]
=2\pi\,
{\cal P}_l(\zeta)\ {\cal P}_l(\zeta')\;.
\label{additionth1}
\ee
This demostrates that the set of polynomials (\ref{newset}) corresponds
to the expansion in partial waves of produced pions.

%
% BIBLIOGRAPHY
%

%

\newpage

\begin{center}
\begin{table}
\begin{tabular}{|c|c|c|c|c|c|c|}
\hline
$(nl)$ & $B_{nl}^\|(0)$ & $B_{nl}^\perp(0)$ & $c_1^{\|(nl)}$~GeV$^{-2}$
& $c_1^{\perp (nl)}$~GeV$^{-2}$&
$\frac{d}{dw^{2}}\log\, B_{nl}^\|(0) $~GeV$^{-2}$    &
$\frac{d}{dw^{2}}\log\, B_{nl}^\perp(0) $~GeV$^{-2}$ \\
\hline
(01)& 1 & 1 & 0 & 0 &1.46 &  0.68      \\
(21)&-0.2& 0.2 & -0.6 & -0.95&0.85&-0.27\\
(23)&0.26& 0.2 & 0&0  &0.65&0.27\\
(41)&-0.006&0.05& 1.1&-0.92 &2.5&-0.24\\
(43)&-0.12 & 0.11&-0.36&-0.28 &0.30&-0.01\\
(45)&0.14& 0.08& 0 & 0 &0.36&0.15\\
\hline
\end{tabular}
\caption{ Values of the subtraction constants entering
eqs.~(\protect{\ref{thesolution}},\protect{\ref{rho2pi}})
 calculated in the model
of instanton vacuum. Isovector case.}
\end{table}

\begin{table}
\begin{tabular}{|c|c|c|c|}
\hline
 & our results & Ball \& Braun & Bakulev \& Mikhailov\\
\hline
$f_\rho^\|$ (MeV)& 190  & $198\pm 7$ & $201\pm 5$ \\
$f_\rho^\perp$ (MeV)& 140  & $160\pm 10$ & $169\pm 5$ \\
\hline
$a_2^\|$ & -0.14  & $0.18\pm 0.10$ & $0.08\pm 0.02$ \\
$a_4^\|$ & -0.01  & -- & $-0.08\pm 0.03$ \\
$a_2^\perp$ & 0.11  & $0.2\pm 0.10$ & $0.36\pm 0.03$ \\
$a_4^\perp$ & 0.03  & -- & -- \\
\hline
\end{tabular}
\caption{Parameters of the twist-2
 vector meson distribution amplitudes
extracted from the $2\pi$DA's compared with
QCD sum rule calculations [34,35].
Our results refer
to normalization point $\mu\approx 600$~MeV, QCD
sum rule calculations to $\mu\approx 1$~GeV. }
\end{table}

\end{center}

\newpage

\end{document}